\documentclass{epl}
\usepackage{graphicx}

\title{Topological defects, pattern evolution, and hysteresis in thin magnetic films}

\author{P. A. Prudkovskii\inst{1}, A. N. Rubtsov\inst{1} and M. I.
Katsnelson\inst{2}}

\institute{\inst{1} Department of Physics, Moscow State University
- 119992 Moscow, Russia\\ \inst{2} Institute for Molecules and
Materials, Radboud University Nijmegen - 6525 ED Nijmegen, The
Netherlands}

 {Domain walls and domain structure}



\begin{document}

\maketitle

\begin{abstract}
Nature of the magnetic hysteresis for thin films is studied by the
Monte-Carlo simulations. It is shown that a reconstruction of the
magnetization pattern with external field occurs via the creation
of vortex-antivortex pairs of a special kind at the boundaries of
stripe domains. It is demonstrated that the symmetry of order
parameter is of primary importance for this problem, in
particular, the in-plane magnetic anisotropy is necessary for the
hysteresis.
\end{abstract}

\section{Introduction}

Traditional views assume that under the thermodynamic equilibrium
conditions a system should either be homogeneous or consist of
macroscopically large domains of homogeneous phases. It appears,
however, that  equilibrium or very long-lived metastable states
occur frequently with a mesoscale heterogeneity (modulated phases,
patterns, etc.; for a general review see, e.g., Ref.\cite{seul}).
These inhomogeneous states may be either regular (e.g., stripes,
stripe domains) or ``chaotic''. The stripes in high-temperature
superconductors and doped Mott insulators \cite{highTc,emery},
supramolecular self-assembly in organic chemistry \cite{chem}, and
``heterogeneous fluctuations'' in metallic alloys \cite{pack}
should be mentioned in this context. Stripe magnetic domain state
of thin ferromagnetic films \cite{kittel} under certain conditions
turns out to be unstable with respect to the formation of
complicated two-dimensional ``chaotic'' pattern
\cite{garel,seul1,brucas}, which provides another example of the
mesoscale pattern evolution, interesting not only conceptually but
also for the applications related to the information storage.

It is commonly accepted now that the formation of the mesoscale
heterogeneity is a result of frustrations in the system which can
result from either geometric factors \cite{pack,kleman,nelson} or
competing interactions, the long-ranged forces such as Coulomb or
dipole-dipole interactions being of primary importance
\cite{emery,garel,kivelson,nussinov,schmalian}. General concepts
of ``avoided criticality'' \cite{kivelson,nussinov} and
``self-induced glassiness'' \cite{schmalian} have been proposed to
treat this situation. However, we still have no detailed theory to
describe a formation of heterogeneous mesoscale states in
frustrated systems. In particular, there is no satisfactory
description of the experimentally observed \cite{seul1,brucas}
``chaotization'' of the magnetic stripe domains beyond the
oversimplified Ising model; for the latter, several Monte-Carlo
\cite{booth,tarjus} or phase-field \cite{jagla} simulations have
been carried out. However, the vector character of the order
parameter is of crucial importance for the magnetism.

In particular, the key issues for the applications such as the
pattern evolution with the in-plane magnetic field variation and
the magnetic hysteresis cannot be formulated in the framework of
the Ising model. These questions are investigated in the present
work.

Stripe domain formation is the result of an interplay of
magnetostatic energy, energy of domain walls, and easy-axis
magnetic anisotropy \cite{kittel}. For the case of magnetic
multilayers, the latter comes mainly from the layer interfaces.
Easy magnetization axis normal to the film plane $xy$ appears
\cite{brucas}. Further we will argue that the in-plane anisotropy
is also crucial for the magnetic behavior of the films. To
demonstrate this we consider two limiting cases, that are the
model with the two-component magnetization vector lying in the
$yz$-plane and the three-component (Heisenberg) case without any
$xy$-anisotropy.

\section{Model and simulation results}

We start with the following effective  Hamiltonian for the system
under consideration:
\begin{eqnarray}\label{H0}
    &H=\int \left( \frac{J_x}{2} \left(\frac{\partial {\bf m}}{\partial x}\right)^2+\frac{J_y}{2}
     \left(\frac{\partial {\bf m}}{\partial y}\right)^2 - \frac{K}{2} m_z^2-h m_y\right)
     d^2r+
\\  \nonumber
    &+\frac{Q^{2}}{2}\int\int m_z({\bf r}) \left(\frac{1}{|{\bf r} - {\bf r'}|}-
    \frac{1}{\sqrt{d^2+({\bf r}-{\bf r'})^2}}\right)
    m_z({\bf r'}) d^2 r d^2 r'
\end{eqnarray}
Here ${\bf m}$ is the magnetization unit vector having either two
or three components, $h$ is an external field, $J_x,J_y,K$ are
exchange and anisotropy parameters, and the last term describes
the long-range dipole-dipole interaction \cite{garel} ($Q$ is the
effective magnetic charge density and $d$ is the film thickness).
Note that stripes appear only if the exchange interaction is
anisotropic; a checkerboard-like structure arises for $J_x=J_y$
\cite{brucas}. With our choice of parameters, stripes are formed
along $y$-axis, that is parallel to the external field. The domain
structure arises for not too strong external fields since at $h\to
\infty$ the magnetization vector is just parallel to the $y$-axis.
We will consider the case of a regular stripe domain structure is
stable at zero external field (which is observed, e.g., for the
permalloy-cobalt multilayers in Ref.\cite{brucas}). In these
experiments an instability of the stripe domain structure with the
external field increase has been observed, which can be described
in terms of the appearance of disclinations at the domain wall
lines \cite{seul1,kashuba}; further a two-dimensional ``chaotic''
magnetization pattern is formed.

In our Metropolis Monte Carlo \cite{MC} simulations, a
discrete-lattice (300 by 300 sites) analogous to the effective
Hamiltonian (\ref{H0}) is studied, in particular, the exchange
terms are approximated by appropriate nearest-neighbor
interactions. We present here the data for the system with $J_x=2,
J_y=4, K=4, Q^2=0.1$ and $d=10$ obtained at several values of the
temperature.  The parameters are chosen to produce the stripe
domains with a width $\Lambda$ of about 10 lattice periods.  In
each simulation, we start with a large (saturating) negative
external field and slowly varied it to a large positive value and
then back.
\begin{figure}
\onefigure[width=8cm]{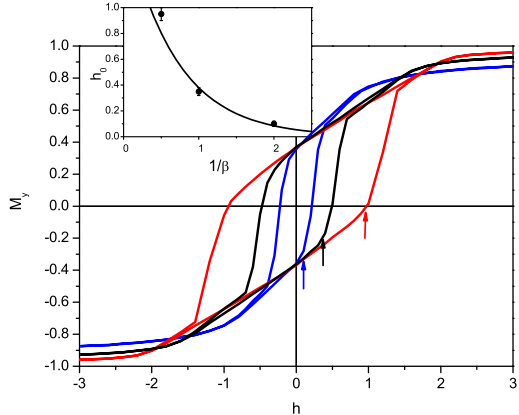}
\caption{ Magnetization curves for the system with two-component
magnetization. Red, black and blue loops correspond to the
temperature values $1/\beta=0.5, 1$ and $2$, respectively; other
parameters are given in the text. Arrows indicate the critical
field, $h$, corresponding to the nucleation of the
vortex-antivortex pairs. The inset shows these values of $h$,
whereas the line in the inset shows the estimation (\ref{h0}) with
$\ln \frac{T L}{t_0 \sigma}=13$}.\label{fig.1}
\end{figure}

\begin{figure}
\onefigure[width=8 cm]{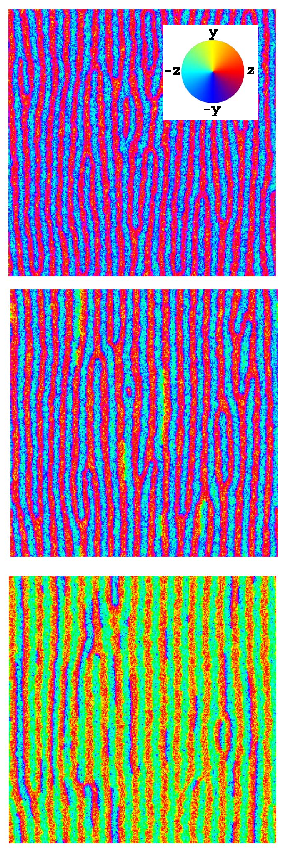} \caption{ Snapshots of the
stripe-domain system with the two-component order parameter at
several points of the hysteresis loop for $\beta=1$. Magnetic
field is $h=0$, $h=0.3$, and $h=0.6$, from top to bottom shot. The
inset shows the color legend for the orientation of local
magnetization.}\label{fig.2}
\end{figure}

Figure \ref{fig.1} shows the magnetization curves $M_{y}(h)$ for
several values of temperature for the two-component ($yz$)
magnetization. Well-pronounced hysteresis loops are clearly seen.
To reveal the source of the hysteresis, we display snapshots of
the system at several points of the hysteresis loop (Figure
\ref{fig.2}). Inside the stripe domains, the magnetization is
oriented almost perpendicular to the film plane, so that $m_z=1$
and $m_z=-1$. These areas are  drawn by red and light-blue colors,
respectively. At a domain wall, the magnetization has to pass
through the points where $m_y=-1$ or $m_y=1$. These points are
shown by and dark-blue and yellow-green colors. Mixtures or the
appropriate colors is used for intermediate orientations of the
magnetization.

At zero field, the magnetization rotates through the negative
$m_y$-direction for all domain walls. At a certain positive $h$,
``green'' parts of the domain walls with positive $m_y$ arise.
Further increase of the external field results in a sudden flip of
all domain walls to positive $m_y$. One can see that the
magnetization vector rotates over $2 \pi$ angle at the motion
along a closed loop containing the end point of the ``green''
part. Thus, the magnetization flip regions of the domain walls are
terminated by vortex-antivortex pairs. It should be stressed that
here both the vortex and antivortex always locate at the same
linear domain wall contrary to the Kosterlitz-Thouless situation
\cite{KT} where the vortices can travel in the entire plane.

\begin{figure}
\onefigure[width=8 cm]{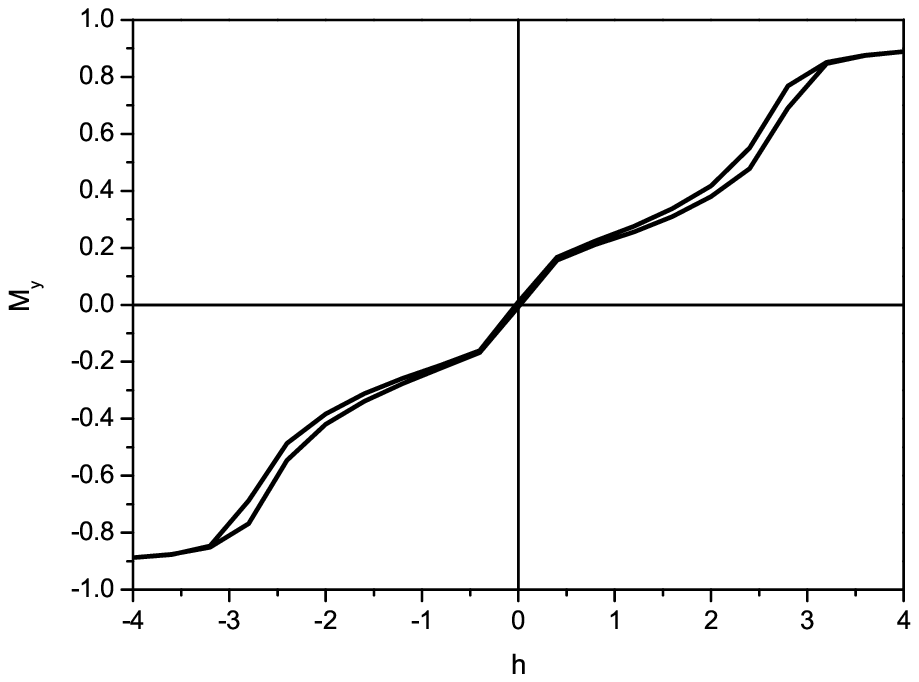} \caption{ Magnetization curve for
the case of tree-component magnetization at temperature
$1/\beta=1$. All the parameters are the same as for $1/\beta=1$
magnetization curve in Figure 1.}\label{fig.3}
\end{figure}

Similar computations have been performed for the case of
three-component magnetization (Heisenberg ferromagnet with
dipole-dipole interactions). The system has a similar stripe
domain structure. However, in the very contrast to the
two-component case, no hysteresis is observed within the errorbar
of calculation, as Figure \ref{fig.3} shows.

\section{Physical picture}

Let us first discuss the source of hysteresis for the
two-component magnetization. Consider the energy balance of a
single vortex-antivortex pair starting with the case $h=0$. First
of all, creation of the vortex itself requires a finite energy.
Near the vortex singularity, energy is mainly contributed by the
exchange (gradient) terms in Eq.(\ref{H0}), similarly to XY-model
\cite{KT}. The size of this area can be estimated as a smaller
value from the vortex-antivortex distance $l$ and domain wall
thickness $\sigma = \sqrt{J_x/K}$. A standard estimation with the
logarithmic accuracy similar to Ref.\cite{KT} results in the
expression for the energy of pair at $h=0$:
\begin{eqnarray}
E(l<\sigma)=2 \pi \sqrt{J_x J_y}\ln \left(l/r\right),\\
\nonumber E(l>\sigma)=2 \pi \sqrt{J_x J_y}\ln
\left(\sigma/r\right)
\end{eqnarray}
Here the short-range cutoff of order of the lattice constant,
$r_0$, is introduced.

The existence of hysteresis means that the value of $E(l>\sigma)$
is large enough in comparison with the temperature, otherwise
thermal fluctuations should destroy oriented domain walls. The
external field makes the wall flip energetically favorable and
decreases the barrier between the the flipped and unflipped wall.
Let us estimate the height of this barrier. We will discuss the
lower part of the magnetization curve, so that positive $h$ tends
to flip the domains. Consider the term $-h m_y$ as a perturbation.
The zero-order approximation for $\bf{m}(\bf{r})$ can be
constructed explicitly for $l\ll\sigma$, because depolarization
effects and the anisotropy term in (\ref{H0}) can be neglected in
this case. The first-order correction to $E(l\ll\sigma)$ appears
to be $h \left(\int m_y d^2 r\right)_{h=0}=-\pi \ln 2 \sqrt{J_x
J_y} l^2 h$. For large $l$, this correction is obviously
proportional to $l$, that is $E(l\gg\sigma)\propto - l h/\sigma$.
One can see that the function $E(l)$ at non-zero positive $h$ has
a maximum at $l\ll\sigma$ or $l\sim\sigma$. Use the limit of small
$l$ for the logarithmic-accuracy estimation. We obtain that the
maximum of $E(l)$ take place at $l_{max}^2\propto J_x/h$ and
$E(l_{max})\approx \pi \sqrt{J_x J_y} \ln \frac{J_x}{r_0^2 h}$.
There is a noticeable probability of the vortex-antivortex pair
nucleation at the domain wall interval of the length $L$ during
the time $T$, if $e^{-\beta E(l_{max})}\approx \frac{T L}{t_0
\sigma }$, where short-time cut-off $t_0$ is the shortest
relaxation time of the system \cite{Kramers}. Therefore we
estimate that the vortex-antivortex pairs arise at
\begin{equation}\label{h0}
    h\propto \frac{J_x}{r_0^2} \exp \left(- \frac{\ln \frac{T L}{t_0 \sigma}}{\pi \beta \sqrt{J_x
    J_y}}\right).
\end{equation}

It is worthwhile to note at this point that, strictly speaking,
the applicability of the Monte Carlo scheme is proven only for the
thermodynamic equilibrium case. However, it is clear that for the
situation of several well-separated local minima, individual
properties of those metastable states are also described
correctly. The transition through energy barrier between the
minima in the Metropolis Monte Carlo simulation is in fact a
diffusion process with a typical timescale $t_0$ of about a single
Monte Carlo sweep, therefore in our numerical calculations $\ln
\frac{T L}{ t_0 \sigma}\approx 10 \div 15$. The obtained
dependence agrees well with the numerical data, as the inset in
Fig. 1 shows.

Possible effects of interaction between domain walls should be
taken into account. To determine this interaction, it is
sufficient to consider only the structure which is homogeneous in
the $y$-direction. One should note first that the domain wall flip
does not affect the dipole-dipole interaction term since this flip
does not change $z$-component of the internal magnetic field.
Therefore we can ignore the last term in Eq.(\ref{H0}) in our
estimation. At zero external field, Eq.(\ref{H0}) reads
\begin{equation}\label{dyn}
    \int \left( \frac{J_x}{2} \dot{\phi}^2 -\frac{K}{2} \sin^2 \phi\right)
    dx=min,
\end{equation}
where the phase $\phi$ describing the orientation of two-component
magnetization is introduced: ${\bf m}=(\sin\phi, 0, \cos \phi)$.
This is equivalent to the minimal action condition for a pendulum,
if $x$ acts as a time \cite{mechanica}. Qualitatively, the domain
wall corresponds to a potential energy minimum of the pendulum
$\phi=0, \pm \pi, ...$, whereas domains themselves are described
by the near-separatrix motion $\phi\approx\pm \pi/2, \pm 3\pi/2,
....$. Domains with parallel and antiparallel walls are described
by the conditions $\phi(0)=0, \phi(\Lambda)=0$ and $\phi(0)=0,
\phi(\Lambda)=\pi$, respectively. The difference between the
values of the action for these two trajectories gives an effective
interaction energy of domain walls. For the case of
$\Lambda/\sigma\gg 1$, the parallel wall configuration is
energetically favorable, but the energy gain is exponentially
small with a factor of $e^{-\Lambda/\sigma}$. In our simulations
as well as in the experiment \cite{brucas} $\Lambda/\sigma\approx
7 \div 10$, and this exponentially small term can be neglected. It
would be interesting to realize experimentally a system with
$\Lambda\approx\sigma$, where the interaction between domain walls
would be important.

In the case of the Heisenberg model (three-dimensional vector
order parameter without $xy$ anisotropy), the vortices and
antivortices will not arise since the orientations of domain walls
can be changed without a transition over an energy barrier. For
example, for the domain profile passing from $m_z=1$ to $m_z=-1$
orientation via $y$-direction, one can consider the 180$^0$
rotation of $\bf m$ in the $xy$-plane. The magnetization
distribution $m_x = \sin \phi \sqrt{1-m_{z}^{2}}, m_y = \cos \phi
\sqrt{1-m_{z}^{2}}$ interpolates smoothly between the domain walls
passed through $+y$ and $-y$ ($\phi = 0$ and $\phi = \pi$) with
the energy independent on $\phi$, as one can see from
Eq.(\ref{H0}). In agreement with these simple topological
considerations, our numerical simulations do not show any
hysteresis effects for the Heisenberg case. This demonstrates the
relevance of in-plane magnetic anisotropy for the hysteresis in
magnetic films and multilayers. This anisotropy can suppress a
continuous rotation of the magnetization vector and therefore
result in a hysteresis behavior.

We investigate here a simplified model of the stripe-domain
formation; in the conclusion, we discuss its relation with real
experimental situation. Although the Hamiltonian (\ref{H0}) is
frequently used to describe the stripe domains \cite{garel}, it
may by a concern that the dipole-dipole interactions due to the
in-plane magnetization components are neglected in this approach.
Indeed, the effective density of magnetic charges is $-Q~{\rm
div}~{\bf m}$, and only the surface charges due to the
discontinuity of the normal component of $\bf{m}$ at the film
boundary are taken into account in Eq.(\ref{H0}). The bulk charge
density is proportional to $\frac{\partial m_x} {\partial
x}+\frac{\partial m_y} {\partial y}$ which is zero inside the
domains but not in the domain walls. Physically, the in-plane
dipole moment of the domain wall may occur. The detailed study of
these effects is out of the scope of the paper; let us only
justify why they can be neglected in the cases under
consideration. First of all, the bulk magnetic charges do {\it
not} appear for the two-component order parameter since $m_x$ is
absent and $m_y$ is independent of $y$. For the three-component
order parameter, the domain wall may have a dipole moment of order
of $Q \sigma d$ per unit wall length. This value should be
compared with similar quantity of the domain, that is $Q \Lambda
d$. Therefore our consideration is valid assuming that $\sigma \ll
\Lambda$. Qualitatively, even in this case the bulk charges
introduce some anisotropy in the $xy$-plane and can therefore
result in hysteresis effects for the three-component
magnetization, but these effects should be much weaker than for
the two-component case.

\acknowledgments

The work was supported by the Netherlands Organization for
Scientific Research (NWO project 047.016.005). One of the authors
(A.R.) is grateful to the "Dynasty" foundation for the support.

\end{document}